\documentclass[letterpaper, times, mathptm,10 pt, conference]{ieeeconf}  

\IEEEoverridecommandlockouts                              

\usepackage{amsmath,amssymb,color,url,graphicx,algorithm,subcaption,bm}
\usepackage{mathtools}
\usepackage{xurl}

\newtheorem{proposition}{Proposition}

\usepackage{algpseudocode}

\title{\LARGE \bf
Integrated equilibrium model for electrified logistics and power systems
}

\author{Rui Yao$^{1}$, Xuhang Liu$^{1}$, Anna Scaglione$^{2}$, Shlomo Bekhor$^{3}$, Kenan Zhang$^{1}$
\thanks{$^{1}$School of Architecture, Civil and Environmental Engineering, École Polytechnique Fédérale de Lausanne (EPFL), Lausanne, Switzerland ({\tt\small{rui.yao, xuhang.liu, kenan.zhang}@epfl.ch}).}
\thanks{$^{2}$Department of Electrical and Computer Engineering, Cornell Tech, New York, United States ({\tt\small as337@cornell.edu}).}
\thanks{$^{3}$Faculty of Civil and Environmental Engineering, Technion--Israel Institute of Technology, Haifa, Israel ({\tt\small sbekhor@technion.ac.il})}
}

\begin{document}

\maketitle
\thispagestyle{empty}
\pagestyle{empty}

\begin{abstract}
This paper proposes an integrated equilibrium model to characterize the complex interactions between electrified logistics systems and electric power delivery systems. The model consists of two major players: an electrified logistics operator (ELO) and a power system operator (PSO). The ELO aims to maximize its profit by strategically scheduling and routing its electric delivery vehicles (e-trucks) for deliveries and charging, in response to the locational marginal price (LMP) set by the PSO. The routing, delivery, and charging behaviors of e-trucks are modeled by a perturbed utility Markov decision process (PU-MDP) while their collective operations are optimized to achieve the ELO's objective by designing rewards in the PU-MDP. 
On the other hand, PSO optimizes the energy price by considering both the spatiotemporal e-truck charging demand and the base electricity load. 
The equilibrium of the integrated system is formulated as a fixed point, proved to exist under mild assumptions, and solved for a case study on the Hawaii network via Anderson's fixed-point acceleration algorithm. 
Along with these numerical results, this paper provides both theoretical insights and practical guidelines to achieve sustainable and efficient operations in modern electrified logistics and power systems. 
\end{abstract}

\section{INTRODUCTION}

The rapid electrification of logistics fleets, driven by global sustainability initiatives and advancements in electric vehicle (EV) technology, is reshaping urban logistics and electric power systems demand. Companies such as Amazon, FedEx, and Shunfeng have pioneered large-scale adoption of electric trucks (e-trucks) for delivery services, offering a pathway to reduce carbon emissions and operational costs~\cite{samet2023ghg}. However, the introduction of e-trucks results in complex interdependence between urban logistics and electric power systems. On the one hand, the spatiotemporal charging demand of e-trucks is largely driven by their daily operations and thus can hardly be predicted in the same way as the base electricity load. On the other hand, the energy price also influence the logistics operator's strategies for delivery and charging. 
Hence, there is an urgent need for an integrated approach to analyze the coupled electrified logistics and power systems. 

A substantial portion of past research has tackled the coupled logistics, or more broadly, the transportation and power system, in an isolated manner. 
On the power side, many studies focus on harnessing the flexibility of vehicles in charging by aggregating their energy demands as a virtual power plant, but often entail oversimplified assumptions on the implications of pricing and incentive schemes on the transportation system (see e.g. \cite{mohanty2022demand} for a review). 
On the transportation side, research has been devoted to solving electric Vehicle Scheduling Problems (eVSPs) that respond to economic incentives from the grid while assuming these price incentives are exogenous parameters (see e.g.
\cite{teoh2022electric,wan2024long,zalesak2021real}).
Studies on the equilibrium and economic incentives in the integrated transportation-power system, that capture the closed-loop interactions in charging, are more relevant to this work~(e.g., \cite{alizadeh2016optimal,alizadeh2018retail,cui2021optimal,guo2021stochastic,lai2022pricing,wang2018coordinated,wei2017network,han2018incentivizing,sonmez2024optimal}). However, they primarily focus on charging of personal EVs rather than e-trucks, which would introduce additional modeling complexity due to logistic operations. In addition, compared to personal EV charging (e.g., home charging), logistic e-truck charging demands would be more responsive to electricity prices (e.g., locational marginal prices, LMPs), making classic inelastic demand forecasts inadequate. 

Motivated by the emerging questions and gap in the literature, this paper develops an integrated equilibrium model that captures the complex interactions between an electrified logistic operator (ELO) and a power system operator (PSO). We consider ELO strategically schedules and routes its e-trucks in response to spatiotemporal electricity prices; meanwhile, PSO determines the electricity generation and prices to satisfy both e-truck charging demands and base loads. In the remainder of this paper, we first present the routing, delivery, and charging model of e-trucks established in the framework of perturbed utility Markov decision processes (PU-MDP), then show how the ELO can design rewards to induce cooperation over e-trucks that maximize its overall profit. Next, we formulate the PSO's DC optimal power flow (DC-OPF) problem integrating ELO's charging demand and establish the overall equilibrium in the integrated system. We further show that, under mild assumptions, the equilibrium always exists. Lastly, we present the main findings from a numerical experiment on Hawaii network and explore the mutual impacts of e-truck charging and electricity prices.

\section{MODEL}
Consider an integrated network $\mathcal{G}$ composed of two mutually exclusive subnetworks: the power grid $\mathcal{G}_P$ and the logistics network $\mathcal{G}_R$, i.e., $\mathcal{G}= \mathcal{G}_P \cup \mathcal{G}_R$. The power grid is defined as $\mathcal{G}_P \coloneqq (\mathcal{V}_{P}, \mathcal{E}_{P})$, where $\mathcal{V}_{P}$ consists of  the set of generator $\mathcal{V}_ {P, G}$ and load buses $\mathcal{V}_ {P, L}$, and $\mathcal{E}_{P}$ is the set of branches. 
The logistics network is defined as $\mathcal{G}_R \coloneqq (\mathcal{V}_{R}, \mathcal{E}_{R})$, where $\mathcal{V}_{R}$ denotes the set of operation zones and $\mathcal{E}_{R}$ represents the connections between adjacent zones. Specifically, a subset of operation zones, denoted by $\mathcal{V}_{R, C} \subseteq \mathcal{V}_{R}$, contain charging stations, while another subset $\mathcal{V}_{R, D} \subseteq \mathcal{V}_{R}$ include delivery destinations. Note that $\mathcal{V}_{R, C}$ and $\mathcal{V}_{R, D}$ are not mutually exclusive, i.e., e-truck can charge or make delivery in the same zone. 
The study horizon is discretized into $T$ time steps with equal duration $\delta$, which yields the set of time steps $\mathcal{T}= \{0, 1, ..., T\}$.

\subsection{E-truck's routing, delivery, and charging problem}\label{sec:PU-MDP}
The ELO operates a fleet of $\mathcal{Q}$ homogeneous e-trucks with battery capacity $r_{\max}$ in the unit of energy consumption per time interval $\delta$. All e-trucks depart from a single depot in $O \in \mathcal{V}_{R}$ with full state-of-charge (SOC) at $t=0$, and return to the same depot at $t=T$. Throughout their operations, e-trucks traverse the logistics network $\mathcal{G}_R$ to make deliveries in zones of $\mathcal{V}_{R, D}$ and recharge at stations in $\mathcal{V}_{R, C}$. 
For simplicity, we assume a movement or delivery per time interval consumes one unit of SOC, and the charging rate $\phi\in \mathbb{Z}_+$ is the same among all charging stations. The set of feasible SOC is defined as $\mathcal{R}=\{0, 1,..., r_{\max}\}$.
In addition, we assume each vehicle can make at most $n_{\max}$ delivery stops after departing from $O$, and define the set of feasible delivery stops as $\mathcal{N}=\{0, 1,..., n_{\max}\}$.
Furthermore, we assume charging is available at the depot, and require all e-trucks to be fully charged (i.e., SOC$=r_{\max}$) at $t=T$.

We first consider a decentralized scenario where each e-truck operates independently to maximize its own expected total utility, subject to rewards designed by the ELO. Accordingly, each e-truck's routing, delivery, and charging behaviors can be modeled as a PU-MDP defined by a tuple $(\mathcal{S}, \mathcal{A}, P, u, F, \gamma)$ with each element specified as follows:

\noindent \textit{1) State $s_t \in \mathcal{S}$}. Each state $s_t = (v_t, r_t, n_t, \tau_t)$ describes, at time step $t$, the vehicle's current location $v_t\in\mathcal{V}_R$, SOC $r_t\in\mathcal{R}$, remaining feasible deliveries $n_t\in\mathcal{N}$, and remaining charging time $\tau_t \in \{0, 1, ..., T-t\}$ due to the previous charging decision.

\noindent \textit{2) Action $a_t \in \mathcal{A}_{s_t} \subseteq \mathcal{A}$}. Given state $s_t \in \mathcal{S}$, each e-truck can take five types of actions: i) remaining idle or charging (\textit{I}); ii) make a delivery (\textit{D}); iii) schedule a charging (\textit{C}); iv) move to an adjacent location (\textit{M}); and v) teleport to the depot and/or charge to full battery (\textit{L}). 
We assume vehicles can only make deliveries in their current zones so the first two types of actions are singleton, whereas the set of feasible charging actions is specified as $C(s_t)= \{(\delta_r, \delta_{\tau})|t+\delta_{\tau}\leq T-1, \delta_{\tau} = \delta_r/\phi\}$, where $\delta_r$ is the charging amount and $\delta_{\tau}$ is the charging time; and a movement action is selected from a set $M(s_t)$ that denotes all adjacent zones to the current location. 
Formally, the set of state-specific feasible actions $\mathcal{A}_{s_t}$ is defined as follows:
\begin{itemize}
    \item If $t = T-1$ (reaching the end of operation), and further
    \begin{itemize}
        \item if $v_t \neq O$ or $v_t = O, r_t< r_{\max}$, then $\mathcal{A}_{s_t} = \{L\}$; 
        \item if $v_t = O, r_t=r_{\max}$, then $\mathcal{A}_{s_t} = \{I\}$.
    \end{itemize}

    \item If $r_t = 0$ (out-of-charge), and further
    \begin{itemize}
        \item if $v_t \in \mathcal{V}_ {R, C}$ (charging available), then $\mathcal{A}_{s_t} = \{I\} \cup C(s_t)$;
        \item otherwise, $\mathcal{A}_{s_t} = \{I\}$. 
    \end{itemize}

    \item If $r_t > 0, n_t =0$ (positive SOC but cannot make any delivery), and further 
    \begin{itemize}
        \item if $v_t \in \mathcal{V}_ {R, C}$ (charging available), then $\mathcal{A}_{s_t} = \{I\} \cup C(s_t)  \cup M(s_t) $;
        \item otherwise, $\mathcal{A}_{s_t} = \{I\} \cup M(s_t)$.
    \end{itemize}

    \item If $r_t > 0, n_t > 0$ (positive SOC and feasible deliveries), and further 
    \begin{itemize}
        \item if $v_t \in \mathcal{V}_{R, C}\cap \mathcal{V}_{R, D}$ (charging and delivery available), then $\mathcal{A}_{s_t} = \{I,D\} \cup C(s_t)  \cup M(s_t) $;
        \item if $v_t \in \mathcal{V}_{R, C}, v_t \notin \mathcal{V}_{R, D}$ (charging available), $\mathcal{A}_{s_t} = \{I\} \cup C(s_t)  \cup M(s_t)$;
        \item if $v_t \in \mathcal{V}_{R, D} , v_t \notin \mathcal{V}_{R, C}$ (delivery available), $\mathcal{A}_{s_t} = \{I,D\} \cup M(s_t)$;
        \item otherwise, $\mathcal{A}_{s_t} = \{I\} \cup M(s_t)$.
    \end{itemize}
    
\end{itemize}

\noindent \textit{3) State transition $P:\mathcal{S} \times \mathcal{A} \rightarrow \mathcal{P}(\mathcal{S})$}. Six types of transition are specified based on current state $s_t$ and action $a_t$:
\begin{itemize}
    \item Start-charging: $s_{t+1} = (v_t, r_t+\delta_r, n_t, \delta_\tau-1)$, if $a_t \in C(s_t)$.
    \item In-charging: $s_{t+1} = (v_t, r_t, n_t, \tau_t-1)$, if $ a_t = I, \tau_t > 0$.
    \item Idle: $s_{t+1} = (v_t, r_t, n_t, \tau_t)$, if $a_t = I, \tau_t = 0$.
    \item Delivery: $s_{t+1} = (v_t, r_t-1, n_t-1, \tau_t)$, if $a_t = D$.
    \item Move: $s_{t+1} = (a_t, r_t-1, n_t, \tau_t)$, if $a_t \in M(s_t)$.
    \item Return to the depot: $s_{t+1} = (O, r_t-1, n_{\max}, \tau_t)$, if $a_t = O$.
    \item Teleport to the deport: $s_{t+1}=(O, r_{\max}, n_{\max}, 0)$, if $a_t=L$.
\end{itemize}
Note that the transition ``return to the depot'' means the e-truck starts another shift of delivery and thus the feasible delivery number is reset to $n_{\max}$.

\noindent \textit{4) Reward $u:\mathcal{S} \times \mathcal{A} \rightarrow \mathbb{R}$}. Non-zero rewards are added to three types of actions:
\begin{itemize}
    \item Delivery: $u(s_t, a_t) = \mu_D(t, v_t)$, if $a_t = D$.
    \item Charging: $u(s_t, a_t) = \mu_C(t, v_t, \delta_r)$, if $a_t \in C(s_t)$.
    \item Teleport to the depot: $u(s_t, a_t) = -\rho < 0$, if $a_t = L$.
\end{itemize}
Here, $\mu_D \in \mathbb{R}^{T\times |\mathcal{V}_{R, D}|}, \mu_C \in \mathbb{R}^{T\times |\mathcal{V}_{R, C}| \times r_{\max}}$ are delivery and charging rewards designed by the ELO, which will be explained in the next section, while $\rho$ is a large penalty imposed to vehicles that fail to reach the termination state. 


\noindent \textit{5) Perturbation function $F=(F_{s})_{s\in\mathcal{S}}$:} A set of state-dependent functions $F_s:\text{int}(\Delta_s)\rightarrow\mathbb{R}$ that are essentially smooth and essentially strictly convex at the interior of probability simplex $\Delta_s$~\cite{Rockafellar+1970}. 

\noindent \textit{6) Discount factor $\gamma$} Set to 1 for simplicity in this paper. 

Given the PU-MDP defined above, a representative e-truck finds the optimal routing and charging strategies, summarized by a policy $\pi: \mathcal{S}\rightarrow\mathcal{P}(\mathcal{A})$, that maximizes the expected accumulated rewards over the operation horizon subject to perturbation $F$. The corresponding optimization problem is given by
\begin{equation}\label{eq:PU_MDP_overall}
    \max_{\pi} V^{\pi}(s_0) \coloneqq \mathbb{E}\left[\sum_{t=0}^{T-1}u(s_t, a_t)-F_{s_t}(\pi(\cdot|s_t)) \Bigg| s_0\right],
\end{equation}
where $V^{\pi}(s)$ denote the value on state $s$ and policy $\pi$; the expectation is taken over all possible trajectories $\{(s_t, a_t)\}_{t\geq 0}$ starting from the initial state $s_0$ and following policy $\pi$.

Let $\pi^*$  be the solution to \eqref{eq:PU_MDP_overall} and $V^* \in \mathbb{R}^{|\mathcal{S}|}$ denote the corresponding optimal value vector. The Bellman optimality condition of  \eqref{eq:PU_MDP_overall} suggests
\begin{equation}
    V^*(s_t) = \max_{\pi(\cdot |s_t)} \sum_{a\in\mathcal{A}_{s_t}} \pi(a_t|s_t) Q^*(s_t, a_t) -F_{s_t}(\pi(\cdot |s_t)),
\end{equation}
where $Q^*(s_t, a_t) = u(s_t, a_t)+ \mathbb{E}_{s'\sim P(|s_t, a_t)}[V^*(s')]$ is the optimal Q-value. Moreover, it is proved in \cite{yao2024markov} that $V^*$ exists and is unique for PU-MDP given a termination state.

When the fleet size $\mathcal{Q}$ is sufficiently large, the aggregate behaviors of the homogeneous e-trucks can be represented as continuous flows, as per the common assumption adopted in the transportation literature~\cite{sheffi1985urban}. 
Accordingly, we define $x^*\in \mathbb{R}^{|\mathcal{S}||\mathcal{A}|}$ as the optimal action flows, where each element  $x^*(a_t|s_t)$ denotes the number of e-trucks (in terms of flows) taking action $a_t$ at state $s_t$ under optimal policy $\pi^*$.
Let $q \in \mathbb{R}^{|\mathcal{S}|}$, where $q(s_0) = \mathcal{Q}$ and otherwise zero. The following proposition derived in~\cite{yao2024PUME} connects action flows $x^*$ with optimal values $V^*$.

\begin{proposition}[Adapted from Prop. 2, Lemma 2 in \cite{yao2024PUME}]~\label{prop:link_flows}
Given the PU-MDP specified in Sec.~\ref{sec:PU-MDP}, the optimal value $V^*$ is a continuously differentiable convex function of rewards $u$. In addition, the optimal action flow $x^*$ is a continuously differentiable function of rewards $u$, and further satisfies
\begin{align}\label{eq:link_flow_expression}
    x^*(u) = q^\top\nabla V^*(u), \quad \forall u.
\end{align}
\end{proposition}

Prop.~\ref{prop:link_flows} enables directly computing the delivery and charging flows using the optimal value $V^*(u)$ at reward $u$. 
Let $U_{D} = \{u(s_t, a_t)| s_t \in \mathcal{S}, a_t = D\}$ and $U_{C} = \{u(s_t, a_t)| s_t \in \mathcal{S}, a_t \in C(s_t) \text{ or } \tau_t \neq 0\}$ denote the sets of delivery and charging rewards, respectively, and define $u_D, u_C$ as the corresponding reward vectors. Then, we have
\begin{itemize}
    \item Optimal delivery action flows $x^*_D \in \mathbb{R}^{|U_{D}|}$
    \begin{equation}\label{eq:PUMDP_x_delivery}
        x^*_D(u) = q^\top\nabla_{u_D} V^*(u)
    \end{equation}
    \item Optimal charging action flows $x^*_C \in \mathbb{R}^{|U_{C}|}$
    \begin{equation}\label{eq:charging_demands}
        x^*_C(u) = q^\top\nabla_{u_C} V^*(u)
    \end{equation}
\end{itemize}

\subsection{ELO's reward design problem}\label{sec:reward_design}

Although e-trucks independently optimize their routing and charging strategies, their aggregate behaviors can be coordinated by properly setting the delivery and charging rewards $\mu_D, \mu_C$. Meanwhile, the ELO can also influence the demand by adjusting the spatiotemporal delivery fees. 
In this paper, we consider customers choose among $K$ delivery time slots based on the delivery fees $p_D \in \mathbb{R}^{K|\mathcal{V}_ {R, D}|}$ and assume the corresponding demand function $D: \mathbb{R}^{K|\mathcal{V}_ {R, D}|} \rightarrow \mathbb{R}^{K|\mathcal{V}_ {R, D}|}$ is invertible.

Besides the e-trucks and customers, the ELO also interacts with the electric power system through charging. Let 
$p_c \in \mathbb{R}^{T|\mathcal{V}_ {P, L}|}$ be the spatiotemporal charging prices and $M \in \{0, 1\}^{T|\mathcal{V}_ {P, L}| \times |U_C|}$ be an incidence matrix that connects load buses and charging actions ($M_{ij}=1$ if $j$th charging action is performed at $i$th load bus, otherwise 0). Similarly, an incidence matrix $N\in \{0, 1\}^{K|\mathcal{V}_ {R, L}| \times |u_D|}$ is introduce to connect delivery demand and delivery actions ($N_{ij}=1$ if $j$th delivery action satisfies $i$th delivery demand, otherwise 0).

To derive the optimal rewards, we first formulate the centralized profit maximization problem of ELO as
\begin{subequations}\label{eq:ELO_full_opt}
    \begin{align}
    \max_{x, p_D} \quad & p_D^\top D(p_D) - p_C^\top M  x_C - \rho \mathbf{1}^\top x_L - H(x)\\
    s.t. \quad& D(p_D) = N x_D \label{eq:market_clearance}\\
    \quad & x \in \Omega,
\end{align}
\end{subequations}
where $\Omega$ denotes the set of all feasible e-truck flows (not necessarily optimal), $x_C, x_D, x_L$ correspond to the charging, delivery, and teleporting flows, and $H(x)$ is the perturbation term that captures other operational costs.

Since the demand function $D$ is assumed to be invertible, the market clearance constraint \eqref{eq:market_clearance} can be rewritten as $p_D = D^{-1}(N x_D)$. Accordingly, Problem~\eqref{eq:ELO_full_opt} is reduced to 
\begin{align}\label{eq:ELO_reduced_opt}
    \max_{x\in \Omega} \quad & D^{-1}(N x_D)^\top Nx_D - p_C^\top M  x_C - \rho \mathbf{1}^\top x_L - H(x).
\end{align}

The following proposition establishes the condition of rewards that ensure the solution to Problem~\eqref{eq:ELO_reduced_opt} corresponds to the optimal action flows of PU-MDP.
\begin{proposition}\label{prop:reward_fixed_point}
    Suppose the perturbation term is given by 
    \begin{align}\label{eq:flow_perturbation_func}
        H(x) = \sum_{s_t \in \mathcal{S}}  \left(\sum_{a_t\in\mathcal{A}_{s_t}}x(a_t|s_t)\right) F_{s_t}\left(\frac{x(\cdot|s_t)}{\sum_{a_t\in\mathcal{A}_{s_t}}x(a_t|s_t)}\right),
    \end{align}
    and the state-specific perturbation function $F_{s_t}, \forall s_t \in \mathcal{S}$ are linearly homogeneous. Then, the optimal rewards $\mu^* =(\mu^*_C, \mu^*_D)$ that induce system optimal e-truck operations must be the solution to the following fixed-point problem:
    \begin{subequations}\label{eq:reward_fixed_point}
        \begin{align}
        &\mu_D^* = N^\top  [\nabla D^{-1}(N x^*_D(\mu^*))   N  x^*_D(\mu^*) + D^{-1}(N x^*_D(\mu^*))]  \\
        & \mu^*_C = -M^\top  p_C,
    \end{align}
    where $x^*_D(\mu^*)$ is the optimal delivery action flows of PU-MDP defined in Eq.~\eqref{eq:PUMDP_x_delivery} at a reward determined by $\mu^*$.
    
    Additionally, if the total revenue $D^{-1}(z) z$ is concave in demand $z$, the optimal rewards $\mu^*$ are unique.
    \end{subequations}
\end{proposition}

Prop.~\ref{prop:reward_fixed_point} implies that we can apply fixed point algorithms on Eq.~\eqref{eq:reward_fixed_point} to solve the optimal rewards. Besides, we only need to solve the delivery rewards as the charging rewards are determined by the charging prices. 
The following proposition further proves that the optimal charging flows $x^*_C$ under the optimal rewards $\mu^*$ are continuous in electricity prices $p_C$. 


\begin{proposition}\label{prop:continuity_optimal_charging_action_flows}
    Given PU-MDP specified in Sec.~\ref{sec:PU-MDP} and optimal rewards specified in Prop.~\ref{prop:reward_fixed_point}, 
    the optimal charging flow $x^*_C$ is a continuous function of charging price $p_C$. 
\end{proposition}

We note that Prop.~\ref{prop:continuity_optimal_charging_action_flows} pave the way to defining and solving the equilibrium in the integrated logistics and power system, which will be detailed in Sec.~\ref{sec:equilibrium}. 

\subsection{PSO's DC optimal power flow problem}

Following the common practice, we consider the PSO solves the DC-OPF to determine the generation plan and set electricity price using LMP. The generation shall serve both base loads $l_0\in \mathbb{R}^{T|\mathcal{V}_{P, L}|}$ and the charging demand of e-trucks, which amount to the total demand at load buses as 
\begin{equation}
    l = l_0 + \phi Mx^*_C(p_C).
\end{equation}


Let $g \in \mathbb{R}^{T|\mathcal{V}_{P,G}|}$ be the generation and $\theta \in \mathbb{R}^{T|\mathcal{V}_P|}$ denote the vector of voltage angle. Following the same time discretization, we have $g_t, l_t, \theta_t$ denote the generation, load, and voltage angle at time step $t$, respectively.
The DC-OPF problem is formulated as 
\begin{subequations}\label{eq:DC_OPF}
\begin{align}
    \min_{g, \; \theta} \quad &  \sum_{t \in \mathcal{T}} \left( g_t^\top {C}_2  g_t + c_1^\top g_t \right) & \\
    \text { s.t. } \; & A B A^\top \theta_t = \left[\begin{array}{ c}
        g_t \\
        -l_t
      \end{array}\right], &&\forall t \in \mathcal{T} \label{eq:OPF_conservation}\\
    & \theta_{1,t} = 0, &&\forall t \in \mathcal{T} \label{eq:OPF_slack_bus}\\
    & \underline{g} \leq g_t \leq \overline{ g}, &&\forall t \in \mathcal{T} \label{eq:OPF_generator_limits} \\
    & \underline{{f}} \leq B A^\top \theta_t \leq \overline{{f}}, &&\forall t\in  \mathcal{T} \label{eq:OPF_line_limits}
\end{align}
\end{subequations}
where the quadratic generation cost is determined by a diagonal matrix ${C}_2 = \text{diag}({c}_2)$ with ${c}_2 \in \mathbb{R}^{|\mathcal{V}_{P,G}|}_{++}$, and $c_1 \in \mathbb{R}^{|\mathcal{V}_{P,G}|}_{++}$;
$A \in \{-1, 0, 1\}^{|\mathcal{V}_P|\times |\mathcal{E}_P|}$ is the bus-branch incidence matrix; $B = \text{diag}(b)$ is the susceptance matrix with $b \in \mathbb{R}^{|\mathcal{E}_{P}|}_{++}$; $\bar{g}, \underline{g}, \underline{f}$ and $\bar{f}$ are the maximum and minimum generation as well as the upper and lower line capacity limits, respectively. As per Constraint~\eqref{eq:OPF_slack_bus}, we select bus 1 (one of the generator buses) as slack bus and set its voltage angle $\theta_{1,t} = 0, \;\forall t \in \mathcal{T}$.


Let $g^*, \theta^*$ denote the optimal primal solutions, and $\lambda^*= (\lambda^*_{g}, \lambda^*_{l}, \lambda^*_{0}), \mu^*=(\mu_+^*, \mu_-^*), \eta^*=(\eta^*_+, \eta^*_-)$ denote the optimal dual solutions to Eq.~\eqref{eq:DC_OPF}. Specifically, dual variables $\lambda$ are associated with Constraints~\eqref{eq:OPF_conservation}-\eqref{eq:OPF_slack_bus}, where $\lambda_{g} \in \mathbb{R}^{T|\mathcal{V}_{P,G}|}, \lambda_{l} \in \mathbb{R}^{T|\mathcal{V}_{P,L}|}, \lambda_{0} \in \mathbb{R}^{T}$ correspond to the generator buses, load buses, and slack bus, respectively, while $\mu, \eta$ are associated with Constraints~\eqref{eq:OPF_generator_limits}-\eqref{eq:OPF_line_limits}, respectively. 
Under the LMP mechanism, the electricity price is set to
\begin{equation}
    p_C  = \lambda^*_{l}.
\end{equation}

Note that the optimal solution to Problem \eqref{eq:DC_OPF} depends on the charging demand $x^*_C$. The next proposition shows that the optimal solution $(g^*, \theta^*, \lambda^*, \mu^*, \eta^*)$, so as LMPs, are continuous in $x^*_C$ under mild assumptions.

\begin{proposition}\label{prop:LMP_continuity}
    Suppose the DC-OPF~\eqref{eq:DC_OPF} is feasible in some open neighborhood $X$ of charging demand $x^*_C$, where the set of active constraints remains the same and are linearly independent. Then, the optimal primal and dual solutions $(g^*, \theta^*, \lambda^*, \mu^*, \eta^*)$ are continuous in $x^*_C \in X$. Accordingly, the LMP  $\lambda_l^*(x^*_C)$ is continuous on $X$. 
\end{proposition}

Here, the condition of linearly independent active constraints is necessary to ensure the existence and uniqueness of dual solutions~\cite{wachsmuth2013licq}.

\subsection{Equilibrium of integrated system}\label{sec:equilibrium}

We are now ready to establish the equilibrium between ELO and PSO as a fixed-point problem as follows:
\begin{align}\label{eq:fixed_point}
    p_C = h(p_C) \equiv \lambda_l^*(x^*_C(p_C)),
\end{align}
where $h: \Pi \rightarrow \Pi$ with $\Pi \subseteq \mathbb{R}^{|\mathcal{V_{P,L}}|T}$.

As per Prop.~\ref{prop:continuity_optimal_charging_action_flows} and Prop.~\ref{prop:LMP_continuity}, $x^*_C$ and $\lambda_l^*$ are both continuous, which leads to the continuity of $h$ and accordingly, the existence of equilibrium by evoking the Brouwer's fixed-point theorem. This result is formally stated in the following proposition. 
\begin{proposition}
    Suppose assumptions in Prop.~\ref{prop:continuity_optimal_charging_action_flows} and Prop.~\ref{prop:LMP_continuity} hold, and the set $\Pi$ is compact. Then, there exists $p^*_C \in \Pi$ such that $p^*_C = h(p^*_C)$.
\end{proposition}

\section{NUMERICAL ANALYSIS}
\subsection{Setting and solution algorithm}

\begin{figure}[H]
    \centering
\includegraphics[width=0.85\linewidth]{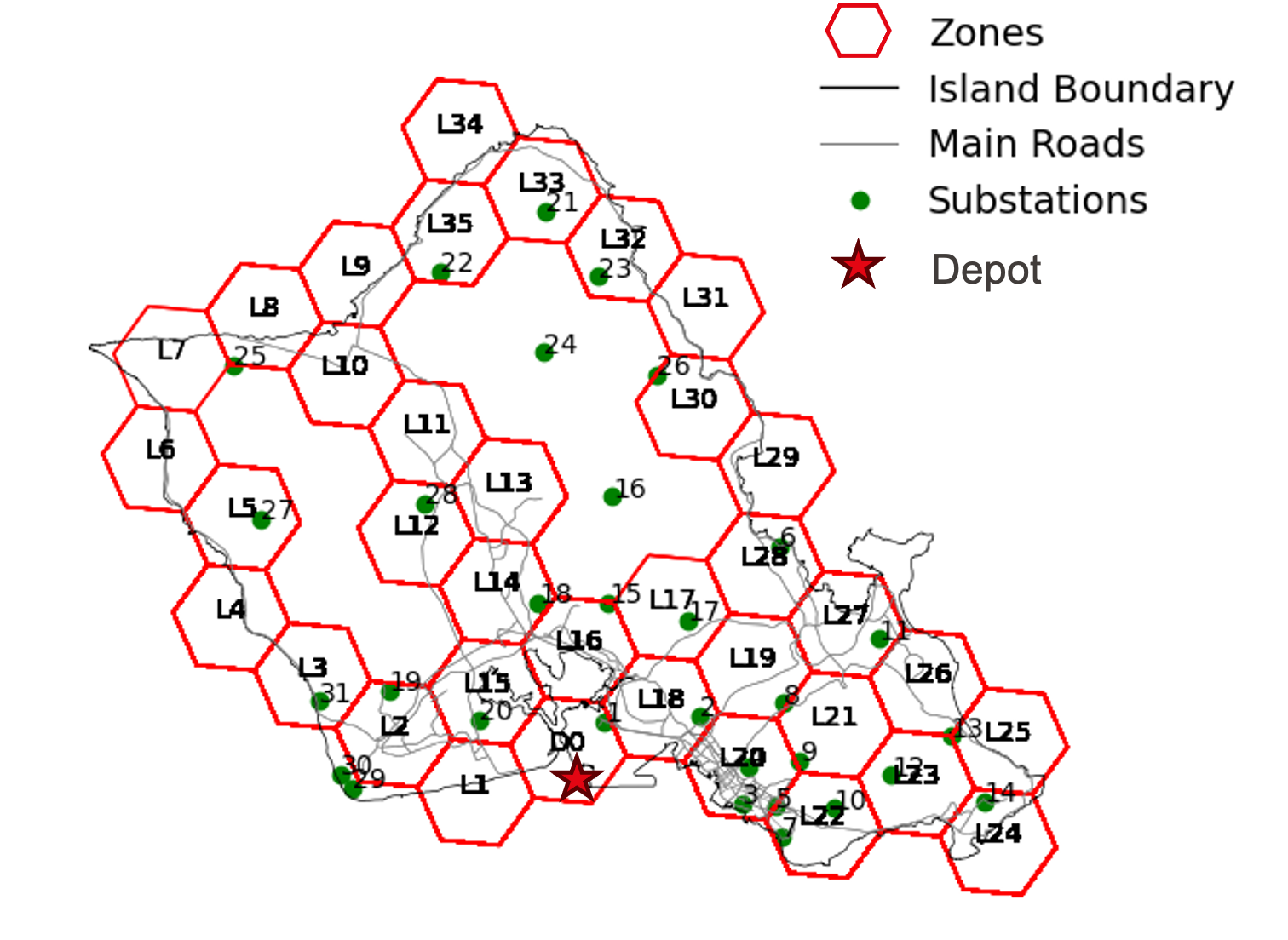}
    \caption{Oahu network (Transportation network: 36 zones, with depot marked as star; Power system: 31 substations, 37 buses, and 45 generators)}\label{fig:network}
\end{figure}

The numerical experiments are conducted on the integrated electrified logistic and power system of Oahu, Hawaii (see Fig.~\ref{fig:network}). The operation horizon ranges from 8:00 to 16:00 with 15-minute intervals. 
The model parameters are summarized in Table~\ref{tab:numerical_parameters}.

\begin{table}[htb]
\centering
\caption{Model Parameters}\label{tab:numerical_parameters}
\begin{tabular}{|c|c|c|}
\hline
\textbf{Notation} & \textbf{Unit} & \textbf{Value} \\
\hline
$\mathcal{Q}$ & veh & 1000 \\
$n_{\max}$ & - & 10 \\
$r_{\max}$ & $\delta$ & 12 \\
$T$ & $\delta$ & 32 \\
$K$ & - & 4 \\
$\delta$ & hr & 0.25 \\
$\phi$ & KW & 150 \\
$\rho$ & - & $10^5$ \\
$|\mathcal{V}_{P,G}|$ & - & 45 \\
$|\mathcal{V}_{P,L}|$ & - & 37 \\
$|\mathcal{V}_{R}|$ & - & 36 \\
\hline
\multicolumn{3}{|c|}{\textbf{Generator parameters by type}}\\
\hline
\text{Generator type} & $c_2$ [$\frac{\$}{(p.u.)^2}$] & $c_1$ [$\frac{\$}{p.u.}$] \\
\hline
Wood & 0.002 & 114.4 \\
Diesel, fuel, oil & 0.004 & 116.5 \\
\hline
\multicolumn{3}{|l|}{\parbox{7cm}{\textbf{Note:} All costs are provided in per-unit (p.u.), and the values of $c_2$ and $c_1$  are the averages each type of generator.}} \\
\hline
\end{tabular}
\end{table}


The perturbation function $F_{s_t}$ is defined as follows:
\begin{equation*}
    F_{s_t}(\pi(\cdot|s_t)) = \pi(\cdot|s_t)^\top [\ln(\pi(\cdot|s_t)) - 1], \forall s_t \in \mathcal{S}.
\end{equation*}

The logistics demand appears in every zone except the depot. Different form e-truck operations, we consider customers choose among two-hour delivery windows, which yields another set of discrete time slots $K$. Following the assumption in Section~\ref{sec:reward_design}, we define the inverse demand function that maps from zonal deliver demand $z$ in time slot $k$ to delivery price $p_D$ as follows:
\begin{equation*}
D_{v, k}^{-1}(z) = 10 - 5 \exp(z/\zeta_v), \forall v \in \mathcal{V}_{R, D}, k \in K,
\end{equation*}
where $\zeta_v$ denote the population at zone $v$~\cite{a2020_census}.



The power system is constructed based on the Hawaii Synthetic Grid~\cite{xu2017application} by aggregating parallel branches. In addition, all charging demands are mapped onto the load buses, so that spatiotemporal LMPs affect charging demands, and vice versa.


\begin{algorithm}[htb]
\caption{Solution algorithm for integrated equilibrium}\label{alg:equilibrium}
\begin{algorithmic}
\State\textbf{Input:} Parameters in Tab.~\ref{tab:numerical_parameters}; PU-MDP $(\mathcal{S}, \mathcal{A}, P, u, F, \gamma)$; network $\mathcal{G}$ with its parameters $M, N, A$, and power system parameters $B, l_0, \underline{g}, \bar{g}, \underline{f}, \bar{f}$. Gap tolerances $\epsilon_1, \epsilon_2$.
\State \textbf{Output}  Equilibrium electricity prices $p^*_C$
\State Initialize $p_C^{(0)}$ as electricity prices without charging.
\State \textbf{for $i=0, 1, ...$ do }
\State \quad \textbf{for $j=0, 1, ...$ do }
    \State  \quad \quad  $\mu_D^{*(j+1)} = \text{AA}_{ELO}(\mu_D^{*(j)}, p_C^{(i)})$ with Eq.~\eqref{eq:reward_fixed_point}
    \State \quad \quad \textbf{If $||\mu^{*(j+1)} - R_{ELO}(\mu^{*(j+1)}, p_C^{(i)})||_2 \leq \epsilon_1$}, stop; \\\quad \quad \textbf{else}, continue.
    \State \quad \quad \textbf{end if}
\State \quad \textbf{end for}
\State \quad Compute charging $x^{*(i)}_C$ under $\mu_D^{*(j+1)},p_C^{(i)}$ with Eq.~\eqref{eq:charging_demands}.
\State \quad Compute LMP $\lambda^*_C(x^{*(i)}_C)$ with DC-OPF~\eqref{eq:DC_OPF}.
\State \quad \textbf{If $||p_C^{(i)} - \lambda^*_C(x^{*(i)}_C)||_2 \leq \epsilon_2$}, stop;  
\State \quad \textbf{else}, $p_C^{(i + 1)} = \text{AA}_{EQN}(p_C^{(i)}, x^{*(i)}_C)$.
\State \quad \textbf{end if}
\State \textbf{end for}
\State \textbf{Return} $p^*_C = p_C^{(i)}$.
\end{algorithmic}
\end{algorithm}

We use value iterations to solve the PU-MDP~\cite{yao2024markov}, and Gurobi 12.0.1 for the DC-OPF problem~\eqref{eq:DC_OPF}. To solve the fixed-point problems for the ELO's reward design and the integrated equilibrium, we employ the Anderson acceleration (AA)~\cite{fu2020anderson}. 
The main idea of AA methods is to construct the next solution $x^{(i+1)}$ by finding the optimal linear combination $\beta^{(i)}$ of past $\Gamma^{(i)}$ iterates that minimizes the norm of fixed-point residual~\cite{fu2020anderson}, i.e., 
\begin{align}
    \min_{\beta^{(i)}} & \sum_{j=0}^{\Gamma^{(i)}-1}{\beta^{(i)}_j} ||x^{(i-\Gamma^{(i)}+j)} - f(x^{(i-\Gamma^{(i)}+j)})||_2, \nonumber \\
     s.t.\;& \sum_{j=0}^{\Gamma^{(i)}-1}\beta_j^{(i)}=1. 
\end{align}
where $f(\cdot)$ is the fixed-point function.

We apply a modified version of the original AA method with additional regularization and safeguarding steps to improve convergence~\cite{fu2020anderson}. We refer to the Appendix for implementation details. 
The solution procedure of the integrated algorithm is summarized in Alg.~\ref{alg:equilibrium}, where $R_{ELO}(\mu^*, p_C)$ denote the RHS of Eq.~\eqref{eq:reward_fixed_point}, and $\text{AA}_{ELO}(\cdot), \text{AA}_{EQN}(\cdot)$ refer to the execution of AA methods on the fixed-point problems for the ELO (Eq.~\eqref{eq:reward_fixed_point}) and the integrated system (Eq.~\eqref{eq:fixed_point}), respectively. 


\subsection{Temporal impacts of charging demand on LMP}\label{sec:results_temporal}

\begin{figure}[htb]
    \centering
\includegraphics[width=1.0\linewidth]{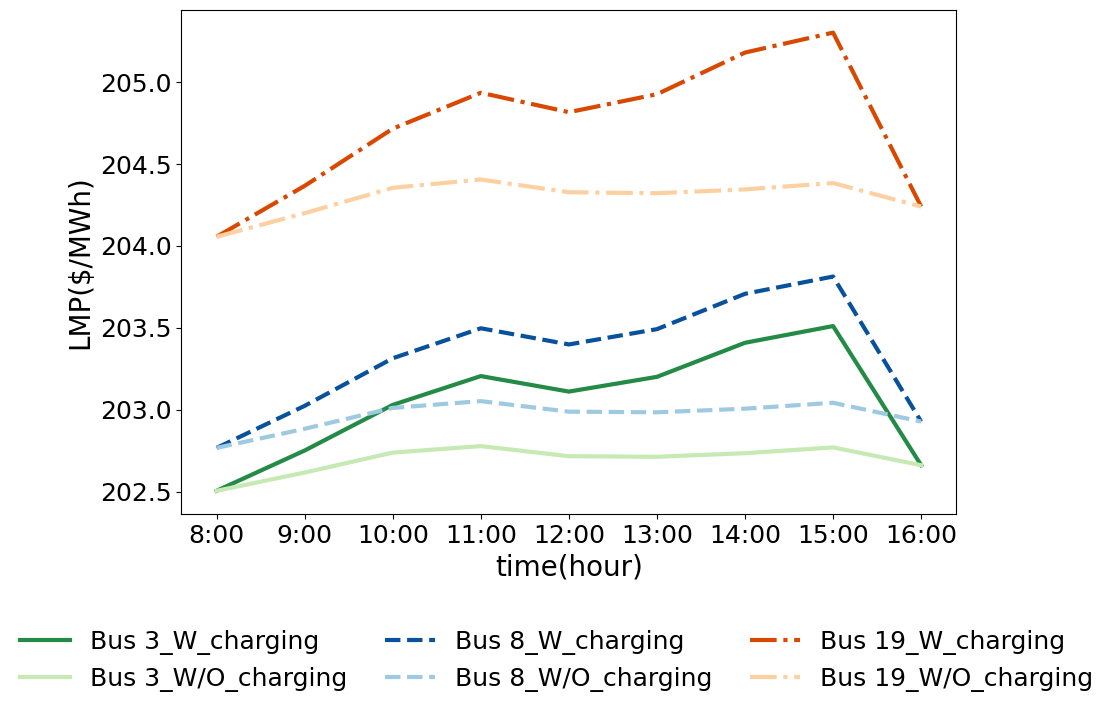} 
    \caption{Impact of charging demand on LMP}\label{fig:charging_impact_on_LMP}
\end{figure}

We first investigate the temporal impacts of charging demand by solving the LMP with and without e-truck operations. 
Fig.~\ref{fig:charging_impact_on_LMP} compares two scenarios of three selected buses (Bus 3, 8, and 19). 
It can be observed that without e-truck charging, LMPs are relatively stable over the study horizon except for a slight increase during 10:00–12:00 ($t=$10-18) and 14:00-15:00 ($t=$24-28). 


As expected, the introduction of e-truck charging induces an increase in LMPs across all three buses after 10:00 ($t=8$), whereas the influence diminishes after 15:00 ($t=28$). This temporal pattern is partly due to the model assumption, as all e-trucks depart from the depot with full battery and return at 16:00. Yet, these results demonstrate that the ELO's operations indeed have impacts on the power system and thus should be well considered in the PSO's decision-making process in anticipation of wider adoption of e-trucks. 



\subsection{Spatiotemporal analysis of LMP with e-truck operations}

\begin{figure}[H]
    \centering
\includegraphics[width=1.0\linewidth]{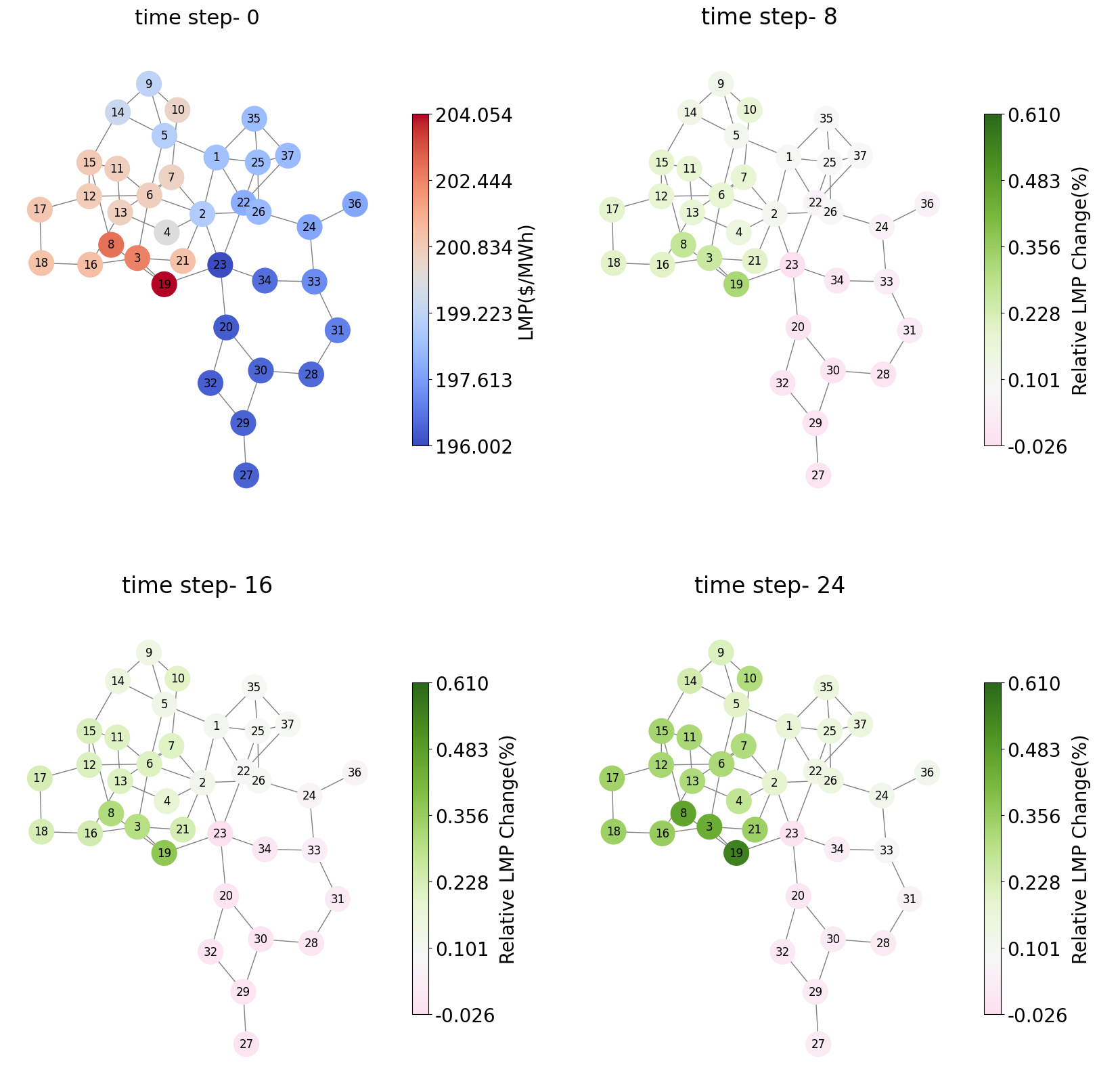} 
    \caption{Spatiotemporal distribution of LMPs}\label{fig:LMP_spatiotemporal}
\end{figure}

Another observation in Fig.~\ref{fig:charging_impact_on_LMP} is that the increase in LMP due to e-truck operations varies among buses. This spatial variation is further illustrated in Fig.~\ref{fig:LMP_spatiotemporal}, which plots the LMP at time $t=0$ and its temporal evolutions. 
As shown in the first subplot, LMPs are much higher at the central-left regions, which correspond to densely populated areas. In particular, Buses, 3, 8, and 19 have the highest LMPs. 
As time proceeds, LMPs in these zones keep increasing and reach the maximum at 14:00 ($t=24$). As discussed in Section~\ref{sec:results_temporal}, such increase is largely contributed by the e-truck charging demand.

Fig.~\ref{fig:LMP_spatiotemporal} also demonstrates that the e-truck charging demand exhibits greater impacts on LMPs when the grid is already congested (e.g., late afternoon and evening peaks), while the impacts during off-peak period are rather subtle.
This finding implies the potential and need for spatiotemporal differential pricing that could further shift charging demands to off-peak hours of the grid.


\subsection{Impact of LMP on e-truck operations}

\begin{figure}[H]
    \centering
\includegraphics[width=1.0\linewidth]{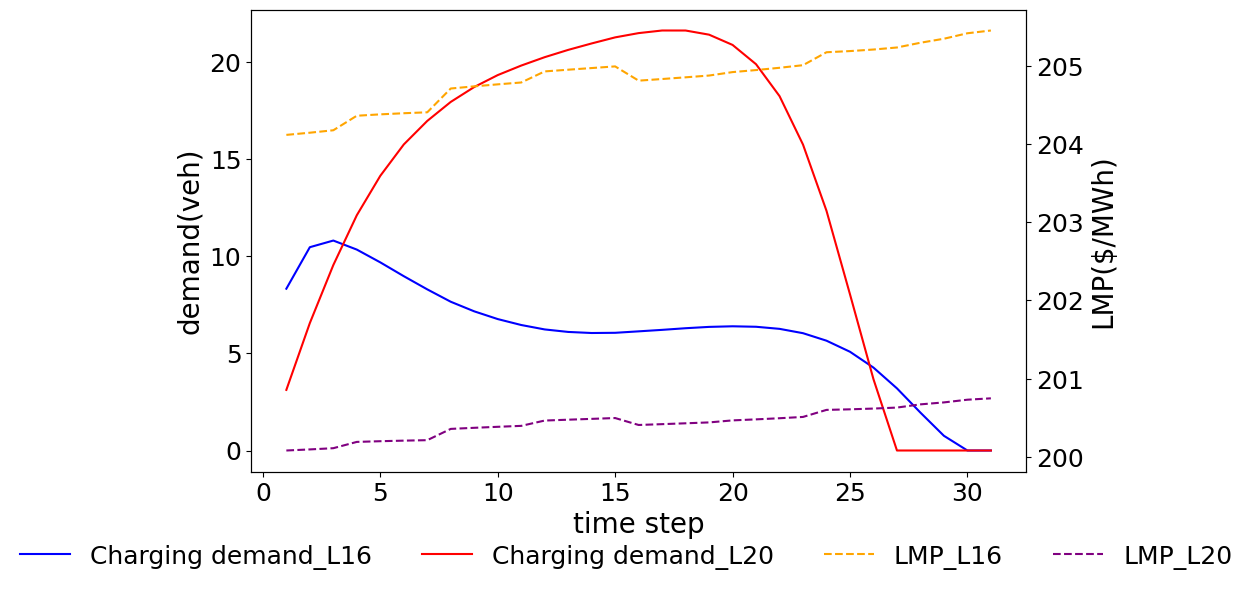} 
    \caption{Impact of LMP on demands}\label{fig:LMP_on_demands}
\end{figure}

We finally analyze how LMP affects the charging behaviors of e-trucks. To this end, we select Zone 16 (adjacent to the depot) and Zone 20 (further away from the depot), and plot the charging demand and LMPs over time (see Fig.~\ref{fig:LMP_on_demands}). 
Since Zone 20 is also far from the central region, its LMP is lower than that in Zone 16 throughout the study horizon. 
Hence, Zone 20 tends to attract more e-trucks to charge, particularly at the end of their operations. 
In contrast, few e-trucks choose to charge in Zone 16 before returning to the depot even though it is closer to the depot. 
This finding confirms the sensitivity of charging demand in response to the spatiotemporal prices. It also indicates that the ELO's operations can be largely affected by the PSO's pricing strategy.

When comparing Figs.~\ref{fig:LMP_spatiotemporal} and \ref{fig:LMP_on_demands}, one can further conclude that the interdependence is not symmetric between ELO and PSO. Since the base loads still contribute to the majority of electricity demand, the influence of e-truck charging is rather minor with up to 0.6\% change in LMP shown in Fig.\ref{fig:LMP_spatiotemporal}. On the other hand, e-trucks' charging strategies are largely driven by LMP, as suggested in Fig.\ref{fig:LMP_on_demands}. Nevertheless, this relationship may shift with the e-truck fleet size and the energy generation.

\section{CONCLUSIONS}
This paper presents an integrated equilibrium model that characterizes the complex interactions between ELO and PSO. A PU-MDP is formulated to capture the decentralized routing, delivery, and charging decisions of individual e-trucks, whose collective behaviors are coordinated via rewards optimized by the ELO.
The ELO and PSO are then coupled through the DC-OPF  problem, where the charging demand of e-trucks influences electricity prices, and vice versa.
We establishes the equilibrium of the integrated system, prove its existence under mild assumptions, and propose an Anderson acceleration-based fixed-point algorithm to solve it. 
The numerical experiments on the real Hawaii network reveal that 
the e-truck charging demand indeed influences LMPs, particularly during peak congestion periods, and the impacts also show spatial variations. Reversely, LMPs largely shape e-trucks' charging decisions and high prices in some zones could greatly suppress the charging demand there. 
In sum, the integrated equilibrium model not only advances our theoretical understanding of the coupled electrified logistics and power system but also provides practical insights into the development of sustainable and resilient urban logistics.

\bibliographystyle{IEEEtran}
\bibliography{references,references2}

\section*{APPENDIX}
\subsection{Proof of Proposition~\ref{prop:reward_fixed_point}}
\begin{proof}
    As per Prop. 3 in~\cite{yao2024PUME}, we have that, if $F_{s_t}(\cdot)$ is linearly homogeneous, there exists an equivalent action flow-based constrained formulation to PU-MDP~\eqref{eq:PU_MDP_overall} as follows, with $H(x)$ specified as in Eq.~\eqref{eq:flow_perturbation_func}:
    \begin{align*}
        \max_{x\in \mathbb{R}_+^{|\mathcal{S}||\mathcal{A}|}} \quad & u^\top x - H(x) \\
    s.t. \quad & (\Lambda - P) x = q, 
    \end{align*}
    where $\Lambda\in \{0,1\}^{|\mathcal{S}||\mathcal{A}|\times |\mathcal{S}|}$, where $\Lambda_{(s_t,a_t), s_t} = 1,\forall s_t\in \mathcal{S}, a_t \in \mathcal{A}_{s_t}$, and zero, otherwise. The corresponding variational inequality (VI) for PU-MDP is to find $x^* \in \Omega$ such that
    \begin{equation*}
        \langle u -  \nabla H (x^*), x^* - x\rangle \geq 0, \;\forall x \in \Omega,
    \end{equation*}
    where $\Omega = \{x| (\Lambda - P) x = q, x\geq 0\}$. Similarly, we have a VI problem for ELO's program~\eqref{eq:ELO_reduced_opt}: to find $x^* \in \Omega^* \subseteq \Omega$, such that
    \begin{equation*}
        \langle \nabla R(x^*) -  \nabla H (x^*), x^* - x\rangle \geq 0, \;\forall x \in \Omega,
    \end{equation*}
    where, we define 
    \begin{align*}
        &\nabla R(x^*) \\&=\begin{cases}
            N^\top  [\nabla D^{-1}(N x_D^*)  N  x_D^* + D^{-1}(N x_D^*)], \text{w.r.t. }x_D^*\\
            -M^\top p_C, \text{w.r.t. }x_C^*\\
             \rho, \text{w.r.t. } {x}_L^*
        \end{cases}
    \end{align*}
    Furthermore, if we have optimal solution to PU-MDP, $x^*$, satisfying
    \begin{align*}
        u - \nabla H (x^*) = \nabla R(x^*) -  \nabla H (x^*),
    \end{align*}
    by examining the two VI problems, we have $x^*$ is also an optimal solution to program~\eqref{eq:ELO_reduced_opt}. 
    In addition, by equivalence to PU-MDP, we have $x^* = x(u^*)$, where $u^*$ is the optimal reward determined by $(\mu_D^*, \mu_C^*)$.
    Hence, the following fixed points characterize the optimal rewards:
    \begin{align*}
        & u^* = \nabla R(x^*(u^*))\\
        \Rightarrow & \mu_D^* = N^\top  [\nabla D^{-1}(N x_D(\mu^*)) N x_D(\mu^*) + D^{-1}(N x_D(\mu^*))]  \\
        & \mu^*_C = -M^\top  p_C.
    \end{align*}
    To prove uniqueness, we first note that concave $D^{-1}(z) z, \forall z \in \mathbb{R}^{|\mathcal{V}_ {R, L}|K}$ implies its Hessian matrix:
    \begin{equation*}
        \nabla^2 D^{-1}(z) z + 2 \nabla D^{-1}(z),
    \end{equation*}
    is negative semi-definite.
    Let $z = N  x_D$, then the Hessian matrix of $D^{-1}(z) z$ with respect to $x_D$ becomes
    \begin{equation*}
        N^\top  [\nabla^2 D^{-1}(z) z + 2 \nabla D^{-1}(z)]  N,
    \end{equation*}
    which is again negative semi-definite. Furthermore, $H(x)$ is strictly convex on polyhedron set $\Omega$~\cite{yao2024markov}. 
    Hence, the objective~\eqref{eq:ELO_reduced_opt} is strictly concave, and $\Omega$ is compact, we have optimal solution $x^*$ exists and is unique.
    
    We now prove uniqueness of $\mu_D^*$ by contradiction: suppose there exists $\mu_{D, 1}^*, \mu_{D, 2}^*$, such that $u_{ 1}^* \neq u_{2}^*$ and $x^*(u_{1}^*) = x^* = x^*(u_{2}^*)$. 
    By the above fixed point, we have
    \begin{align*}
        \mu_{D, 1}^* &=\nabla_{x_D} R(x^*(u_{1}^*)) +  \nabla H(x^*(u_{1}^*)) \\
        &=\nabla_{y_D} R(x_D^*) +  \nabla H(x_D^*) \\
        &=  \nabla_{x_D} R(x^*(u_{2}^*)) +  \nabla H(x^*(u_{2}^*)) = \mu_{D, 2}^*,
    \end{align*}
    which contradicts our assumption. This means fixed point $\mu^*$ is indeed unique.
\end{proof}

\subsection{Proof of Proposition~\ref{prop:continuity_optimal_charging_action_flows}}

\begin{proof}
    We start by defining the convex conjugate function $W(-p_C^\top M) \equiv \max_{x \in \Omega} {D^{-1}(N x_D)}^\top N x_D - p_C^\top M  x_C + \rho^\top x_L- H(x)$. Since $H(x)$ is strictly convex on $\Omega$, by Theorem 11.13 in~\cite{rockafellar2009variational}, convex conjugate function $W: X \rightarrow \mathbb{R}$ is continuous differentiable, where
    $X = \text{int}(\text{dom}(W))$ is not empty and $\text{dom}(W) = \{z\in\mathbb{R}^{|u_C|}: W(z)<+\infty\}$ (i.e., where $W(z)$ is bounded). 
    On the other hand, by Prop.~\ref{prop:link_flows}, $V^*$ exists and is bounded for $u \in \mathbb{R}^{|u_C|}$. As per Proposition~\ref{prop:reward_fixed_point}, optimal solutions of ELO's problem are the same as the PU-MDP under optimal rewards $u^*$, that is $x^*=x(u^*)$. Hence, their objective values are also the same, $W(-p_C^\top M) = V^*(s_0)$, which is bounded. This implies $X = \mathbb{R}^{|u_C|}$.
    Furthermore, since optimal solution $x^*$ is unique (by uniqueness of $\mu^*$ as in Prop.~\ref{prop:reward_fixed_point}), we have that, by continuous differentiability of $W$ on $\mathbb{R}^{|u_C|}$,
    \begin{align*}
        \nabla W(z) &= x^*_C,
    \end{align*}
    which implies $x^*_C(z)\equiv \nabla W(z)$ is also continuous on $\mathbb{R}^{|u_C|}$. In addition, since $z(p_C) = -p_C^\top M$ is continuous and $z: \mathbb{R}^{|\mathcal{V}_ {P, L}|T} \rightarrow \mathbb{R}^{|u_C|}$, we have $x^*_C(p_C)\equiv x^*_C(z(p_C))$ is continuous on $\mathbb{R}^{|\mathcal{V}_ {P, L}|T}$.
\end{proof}

\subsection{Proof of Proposition~\ref{prop:LMP_continuity}}
\begin{proof}
    Under our assumptions, feasible solutions to program~\eqref{eq:DC_OPF} exist.
    Since all constraints in Eq.~\eqref{eq:DC_OPF} are linear, the KKT conditions hold at an optimal solution $(g^*, \theta^*, \lambda^*, \mu^*, \eta^*)$. We have the following KKT system for each $t\in\mathcal{T}$:
    \begin{align*}
        &J_t \cdot \left[\begin{array}{c}
        g_t^*  \\
        \theta_t^*  \\
        \lambda_{0,t}^*  \\
        \lambda_t^*  \\
        \tilde{\mu_t}^* \\
        \tilde{\eta_t}^*  \\
      \end{array}\right] = 
    \left[\begin{array}{c}
        c_1  \\
        0  \\
        0\\
        l_t  \\
        \tilde{g}_t \\
        \tilde{f}_t  \\
      \end{array}\right], J_t\coloneqq 
      \left[\begin{array}{ccc}
        J_t(1) & J_t(2)^\top & J_t(3)^\top  \\
         J_t(2) & \mathbf{0} & \mathbf{0} \\
        J_t(3) & \mathbf{0}& \mathbf{0}  \\
      \end{array}\right] \\
     &J_t(1)\coloneqq 
      \left[\begin{array}{cc}
        2C_2 & \mathbf{0}  \\
         \mathbf{0} & \mathbf{0} \\
      \end{array}\right], J_t(2)\coloneqq 
      \left[\begin{array}{cc}
        \mathbf{0} & e_1  \\
         -\mathbb{I} & Z \\
      \end{array}\right], \\
      &J_t(3)\coloneqq 
      \left[\begin{array}{cc}
        \Phi_g^\top & \mathbf{0}  \\
         \mathbf{0} & \Phi_\theta^\top \\
      \end{array}\right]
    \end{align*}
where $\Phi_{g_t}$, $\Phi_{\theta_t}$ are the Jacobian of the binding inequality constraints~\eqref{eq:OPF_generator_limits}, ~\eqref{eq:OPF_line_limits} (if any) with respect to $g_t$ and $\theta_t$, respectively; and $\tilde{\mu_t}^*, \tilde{\eta_t}^*$ are the corresponding dual variables for the binding inequality constraints, and $\tilde{g}_t, \tilde{f}_t$ are the (either upper or lower) bounds for the binding inequality constraints,  $Z = A B A^\top$, and $e_1 \in \{0, 1\}^{|\mathcal{V}_{P}|}$ is a basis vector at the slack bus. By implicit function theorem, $(g^*, \theta^*, \lambda^*, \mu^*, \eta^*)$ is continuous on an open neighborhood of $l_t$ if $J_t$ is invertible.

Recall that $J_t$ being invertible is equivalent to requiring that $z= 0$ is the unique solution to the following
\begin{align}\label{eq:linear_ind}
    J_t z = 0.
\end{align}

Let $z = (v, w)$, where $v, w$ correspond to the primal and dual variables, respectively. We have
\begin{align}
    J_t(2) v = 0.
\end{align}
Since $Z$ is the weighted graph Laplacian matrix with null space $\text{Null}(Z)= \{\theta|Z \theta = 0 \}=\alpha \mathbf{1}, \alpha \in \mathbb{R}$, the basis vector $e_1$ is independent from $Z$, so that $J_t(2)$ has full column rank. Hence, we have $v = (J_t(2)^\top J_t(2))^{-1} 0 = 0$ and is the unique solution.

Under our assumption, active (i.e., equality and binding inequality) constraints are linearly independent. Hence, $J_t(2,3) \coloneqq [J_t(2)^\top, J_t(3)^\top]$ also has full column rank. We have that $w = (J_t(2,3)^\top J_t(2,3))^{-1} 0 = 0$, which is also unique. We conclude that $J_t$ is invertible, and $(g^*, \theta^*, \lambda^*, \mu^*, \eta^*)$ is continuous on an open neighborhood of $l_t$ (which includes $x_C^*$).
\end{proof}

\subsection{Anderson acceleration (AA) fixed-point iterates}

\begin{algorithm}[H]
\caption{AA fixed-point algorithm}\label{alg:AAFP}
\begin{algorithmic}
\State \textbf{Input} Fixed-point function $f$, regularization parameter $r_{\text{AA}}$, safeguarding parameters $D_{AA} > 0$, $\epsilon_{AA} > 0$, $R_{check} \in \mathbb{Z}_{++}$, max-memory $M_{AA} \in \mathbb{Z}_{+}$, relaxation parameter $\beta_{AA}\in (0, 1]$, stopping criteria $\epsilon_{tol}$.
\State \textbf{Output}  Fixed-point $x^*$.
\State Initialize $n_{AA}=0, R_{AA}=0, I_{init} = \text{True}$.
\State Compute $G^{(0)}=x^{(0)} - f(x^{(0)}), x^{(1)} = x^{(0)} - \beta_{AA} G^{(0)}$
\State \textbf{for $i=1, ...$ do } \Comment{\textit{Note: $\text{AA}_f(\cdot)$ function returns $x^{(i+1)}$}}
\State \quad Choose memory $m_i = \min \{M_{AA}, i\}$
\State \quad Compute gap $G^{(i)}=x^{(i)} - f(x^{(i)})$
\State \quad \textbf{Terminate if $||G^{(i)}||_2 \leq \epsilon_{tol}$.}
\State \quad Compute fallback iterate $\tilde{x}^{(i+1)}=x^{(i)}-\beta_{AA} G^{(i)}$
\State \quad Compute $y^{(i-1)} = G^{(i)} - G^{(i-1)}, s^{(i-1)} = x^{(i)} - x^{(i-1)}$
\State \quad Update memories $Y^{(i)} = [y^{i-m_i}, ..., y^{i-1}],$ and $ S^{(i)} = [s^{i-m_i}, ..., s^{i-1}]$.
\State \quad Solve regularized least square
\begin{equation*}
    \min_{\gamma^{(i)}} ||g^{(i)} - Y^{(i)}\gamma^{(i)}||_2^2 + r_{AA} (||Y^{(i)}||_F^2+||S^{(i)}||_F^2)||\gamma^{(i)}||_2^2
\end{equation*}
\State \quad Compute AA iterate ${x}_{AA}^{(i+1)}=\sum_{j=0}^{m_i} \alpha^{(i)}_j \tilde{x}^{(i-m_i+j+1)}$, where $\alpha^{(i)}_0 = \gamma^{(i)}_0, \alpha^{(i)}_j=\gamma^{(i)}_j - \gamma^{(i)}_{j-1}, \alpha^{(i)}_{m_i} = 1- \gamma^{(i)}_{m_i-1}$.
\State \quad \textbf{if $I_{init}$ or $R_{AA} \geq R_{check}$}:
\State \quad \quad \textbf{if $||g^{(i)}||_2^2 \leq D_{AA}||g^{(0)}||(n_{AA}/R_{check} + 1)^{-1-\epsilon_{AA}}$}:
\State \quad \quad $x^{(i+1)} = {x}_{AA}^{(i+1)}, n_{AA} += 1, R_{AA} = 1, I_{init} = \text{False}$
\State \quad \quad \textbf{else $x^{(i+1)} = \tilde{x}^{(i+1)}, R_{AA} = 0$}
\State \quad \quad \textbf{end if}
\State \quad \textbf{else $x^{(i+1)} = {x}_{AA}^{(i+1)}, n_{AA} += 1, R_{AA} += 1$}
\State \quad  \textbf{end if}
\State \textbf{end for}
\State \textbf{Return} $x^* = x^{(i)}$.
\end{algorithmic}
\end{algorithm}

Our fixed-point iterations adapts from~\cite{fu2020anderson}. In this paper, we define $AA_{ELO}$ by setting fixed-point function $f(\cdot) = R_{ELO}(\cdot)$, and $AA_{EQN}$ with $f(\cdot) = h(\cdot)$. For ELO problem, we set $r_{AA} = 10^{-8}, D_{AA}=10^{5}, \epsilon_{AA}=10^{-5}, R_{check}=10, M_{AA}=5, \beta_{AA}=1, \epsilon_{tol} = 10^{-6}$. For integrated equilibrium, we set $r_{AA} = 10^{-7}, D_{AA}=10^{4}, \epsilon_{AA}=10^{-5}, R_{check}=5, M_{AA}=10, \beta_{AA}=0.1, \epsilon_{tol} = 10^{-4}$.

\addtolength{\textheight}{-12cm}   








\end{document}